\documentclass[11pt]{article}
\usepackage{amsfonts,amsmath,amssymb,bbm}
\usepackage{cite,graphicx,color}
\usepackage[colorlinks=true, linkcolor=blue, citecolor=blue, linktoc=all]{hyperref}
\usepackage[usenames,dvipsnames]{xcolor}
\usepackage{diagbox}
\usepackage{extarrows}
\usepackage{url}
\usepackage[font=small,labelfont=bf,tableposition=top]{caption}
\usepackage{graphicx,esint} 
\usepackage{psfrag}
\usepackage{subfigure}
\usepackage{tabularx}
\usepackage{footmisc}

\newcommand{\ZZ}{\mathbb{Z}}
\newcommand{\RR}{\mathbb{R}}

\newcommand{\myfig}[3]{
	\begin{figure}[ht]
	\centering
	\includegraphics[width=#2cm]{#1}\caption{#3}\label{fig:#1}
	\end{figure}
	}

\newcommand\cc[1]{#1^{^{\kern-6pt \circ}}\kern2pt}

\def\pa{\partial}
\renewcommand{\a}{\alpha}

\newcommand{\m}{\mu}
\newcommand{\n}{\nu}

\def\be{\begin{equation}}
\def\ee{\end{equation}}
\def\bea{\begin{eqnarray}}
\def\eea{\end{eqnarray}}
\def\ba{\begin{array}}
\def\ea{\end{array}}
\def\bi{\begin{itemize}}
\def\ei{\end{itemize}}

\newcommand{\beq}{\begin{equation}}
\newcommand{\eeq}{\end{equation}}
\newcommand{\beqn}{\begin{eqnarray}}
\newcommand{\eeqn}{\end{eqnarray}}
\newcommand{\bga}{\begin{align}}

\def\dalemb#1#2{
	{\vbox{
		\hrule height .#2pt
		\hbox{\vrule width.#2pt height#1pt \kern#1pt\vrule width.#2pt}
		\hrule height.#2pt}}
	}

\DeclareCaptionLabelFormat{andtable}{#1~#2  \&  \tablename~\thetable}

\newcommand{\thistitle}{Aspects of Holography of Taub-NUT-AdS$_4$}
\newcommand{\auth}{
	Institute of Theoretical Physics, Aristotle University of Thessaloniki, 
	54124 Thessaloniki, Greece.
	}
\newcommand{\uiuc}[1]{
	\centerline{
		\begin{minipage}[c]{0.7\linewidth}
			\begin{center}
			${}^{#1}$Illinois Center for Advanced Studies of the Universe \& Department of Physics,\\ 
			University of Illinois, 1110 West Green St., Urbana IL 61801, U.S.A.
			\end{center}
		\end{minipage}
		}
	}

\begin{document}

\allowdisplaybreaks
\title{\thistitle}
\author{
	Georgios Kalamakis,$^{a}$ Robert G. Leigh,$^{b}$ and Anastasios C. Petkou\,$^{a}$
	\\
	\\
	{\small ${}^a$\emph{\auth}}\\ 
	{\small \emph{\uiuc{b}}}
	\\
	}
\date{\today}
\maketitle
\vspace{-5ex}
\begin{abstract}
\vspace{0.4cm}
In this paper we consider aspects of the holographic interpretation of Taub-NUT-AdS$_4$. We review our earlier results which show that TNAdS$_4$ gives rise to a holographic three-dimensional conformal fluid having constant vorticity. We then study the holographic relevance of the Misner string by considering bulk scalar fluctuations. The scalar fluctuations organize naturally  
into representations of the $SU(2)\times \mathbb{R}$ isometry algebra.  If we require the string's invisibility we obtain a Dirac-like quantization relating the frequency of the scalar field modes to the NUT charge. As the latter quantity determines the total vorticity flux of the boundary fluid, we argue that such an assumption allows for a holographic interpretation of TNAdS$_4$ as a {\it non-dissipative} superfluid whose excitations are quantized vortices.  Alternatively, if we regard the Misner string as a physical object, as has recently been advocated for thermodynamically, the aforementioned quantization conditions are removed, and we find that TNAdS$_4$ corresponds to a holographic fluid whose dissipative properties are probed as usual by the complex quasinormal modes of the bulk fluctuations. 
We show that such quasinormal modes are, perhaps surprisingly, organized  into infinite-dimensional non-unitary representations of the isometry algebra.
\end{abstract}

\setcounter{footnote}{0}
\renewcommand{\thefootnote}{\arabic{footnote}}
\newpage

\section{Introduction}
There is by now a considerable amount of evidence that asymptotically locally AdS$_4$ spacetimes are related to three-dimensional conformal fluids in local thermal equilibrium \cite{Rangamani:2009xk}. In particular, exact vacuum solutions of the four-dimensional Einstein equations with a negative cosmological constant determine and are determined by\footnote{The complete classification of the holographic fluids that are dual to four-dimensional Einstein spaces is still an open question and it is related to issues such as black hole uniqueness and rigidity theorems.  See \cite{Gath:2015nxa} for some recent progress.}  a conserved, symmetric and traceless  energy momentum tensor of a three-dimensional fluid and by the background on which the latter resides. In \cite{Mukhopadhyay:2013gja} it was shown that a large class of bulk geometries gives rise to {\it perfect holographic fluids}, namely fluids in {\it global thermal equilibrium} where dissipative effects and hence entropy production are absent. Notably, such {\it perfect bulk geometries} can be explicitly reconstructed in closed form starting from the boundary fluid data, which appears to point towards an underlying integrability of the gravitational systems \cite{Gath:2015nxa,Petropoulos:2015fba}. 

In perfect holographic fluids one can clearly identify their globally defined hydrodynamic variables, such as the temperature, energy density and pressure, which satisfy the usual thermodynamic relations. Moreover, it was shown in \cite{Mukhopadhyay:2013gja} that a crucial requirement of holographic perfect fluidity is the absence of shear in the kinematics of the boundary fluid, nonetheless non-trivial flows are also allowed as the boundary fluids can still have non-zero vorticity. A notable example is the rotating holographic perfect fluid dual to Kerr-AdS$_4$ (KAdS$_4$) spacetimes \cite{Caldarelli:1999xj}. Note that the holographic interpretation of Kerr-AdS took actually some time to be settled \cite{Gibbons:2004ai}. 

Another class of  perfect bulk geometries are Taub-NUT-AdS (TNAdS) spacetimes \cite{Chamblin:1998pz}. The thermodynamic interpretation of Taub-NUT (TN) geometries, which involves studying  analytically continued Euclidean versions, has been a work in progress for a considerable time \cite{Clarkson:2002uj,Astefanesei:2004kn,Johnson:2014xza}. In fact the details depend on the treatment of the Misner string. Quite recently  the Lorentzian versions of Taub-NUT spacetimes were critically revisited \cite{Kubiznak:2019yiu,Bordo:2019tyh}. These results are based on the observation that Lorentzian TN spacetimes  can be "rehabilitated"  even in the presence of Misner strings as they are geodesically complete, i.e. free-falling observers do not "see" the Misner string \cite{Clement:2015aka, Clement:2015cxa}.  It has been claimed that they can be given a consistent thermodynamic interpretation, containing a First Law with an independently varied NUT charge, without imposing time periodicity to avoid the Misner string.

In the context of holography, TNAdS spacetimes were studied in \cite{Leigh:2011au,Leigh:2012jv} where it was shown that they give rise to holographic perfect fluids with constant vorticity, that is, in a vortex flow.  To study the hydrodynamic properties of such a fluid, one needs to perturb it, which amounts to studying the quasi-normal modes (QNMs) for scalar, vector or tensor perturbations around the bulk background (see \cite{Birmingham:2001pj} and \cite{Kovtun:2005ev,Berti:2009kk,Konoplya:2011qq} for reviews). Equivalently, one may consider directly the hydrodynamic fluctuations in the boundary fluid (e.g., \cite{Romatschke:2009im}). Although the literature on holographic hydrodynamics is already enormous and has given very interesting results in the context of AdS/CMT (see e.g., \cite{Hartnoll:2016apf} and references therein), holographic systems in non-trivial background geometries are much less explored. Indeed, despite interesting recent works dealing with the holographic effects of bulk symmetry breaking e.g. \cite{Donos:2019txg},  not much is known about the transport properties of holographic fluids with nonzero vorticity. For example, after the first holographic analysis of QNMs for the global Schwarzchild AdS$_4$ black hole in  \cite{Michalogiorgakis:2006jc}, it took some time until the analogous discussion was presented for the Kerr-AdS$_4$ metric and its boundary fluid in \cite{Uchikata:2009zz,Cardoso:2013pza}.  Another interesting work in that direction is  \cite{Eling:2013sna}. Quite generally, it is a hard problem in black hole physics to find stable solutions that can accommodate asymptotically non-trivial rotational dynamics \cite{Markeviciute:2017jcp}. 

In this work we begin with a review of our earlier results  \cite{Leigh:2011au,Leigh:2012jv} regarding the kinematics of the boundary TNAdS$_4$ fluid. We point out that an important quantity is the integral of the fluid's vorticity over the total fluid surface. This quantity, which may be termed {\it total circulation} or {\it total vorticity flux} (TVF),  makes sense when the fluid resides on a compact two-dimensional spatial surface. Unsurprisingly, the TVF of the TNAdS$_4$ fluid is non-zero and proportional to the NUT charge. We contrast this  with the corresponding result for the holographic Kerr-AdS$_4$ fluid whose TVF is zero. We conclude that the NUT charge is intimately related to the different global rotational properties of the TNAdS$_4$ and KAdS$_4$ fluids, both of which are otherwise locally rotating. 

Next, we study probe scalar fluctuations in a fixed TNAdS$_4$ background. Such a calculation is relevant for studying the possible hydrodynamic properties of the boundary fluid as they may be regarded (for suitable bulk mass) as part of the general metric fluctuations (which we will consider in a separate publication). As well, scalar fluctuations on TNAdS$_4$ are interesting in themselves because they represent the simplest possible system and yet display a number of issues that we must resolve if we are to come to an understanding of holography in this background. 
In particular, we face the question of the holographic relevance of the Misner string.  We will show that we have two options. If we require the invisibility of the Misner string, which leads to the presence of closed time-like curves, we obtain  a Dirac-like quantization that relates the scalar field modes with the NUT charge. However, the NUT charge is proportional to the TVF of the boundary fluid, hence we may interpret these scalar modes as quantized rotating modes in the boundary. 

On the other hand, if we consider the Misner string as a physical object, we should excise a point on the fluid's  surface. This follows from the fact that the angular equation for the scalar field fluctuations has a complete set of eigenfunctions only under such circumstances. Clearly, this is the point where the Misner string touches the boundary. As in Refs. \cite{Leigh:2011au,Leigh:2012jv}, we  interpret this point as the location of an anyonic quasiparticle. We show that under these circumstances, TNAdS$_4$ gives rise to a holographic fluid whose dissipative properties are probed by the usual quasinormal modes of the scalar fluctuations. 
In this case, we will show that scalar modes satisfying infalling boundary conditions at the black hole horizon are quasi-normal modes with complex frequencies, and that these modes fall into infinite-dimensional highest- and lowest-weight representations of the $SU(2)\times\mathbb{R}$ isometry algebra, in keeping with the fact that in these circumstances no quantization condition can be consistently imposed. That is, the isometry algebra is represented non-unitarily and the scalar modes are generically aperiodic, possessing anyonic phases.

The paper is organized as follows. In Section \ref{sec:kinematics}, we review the kinematics of the TNAdS$_4$ and KAdS$_4$ fluids and point out their similarities and differences. In the case of TNAdS$_4$, we emphasize that expected fluid characteristics depend on our treatment of the Misner string, in particular whether it is invisible or physical. Section \ref{sec:scalars} is devoted to studying how scalar field fluctuations behave in each of these situations. After some initial analysis of the scalar system in Section \ref{sec:scalarsetup}, including the isometry algebra, we consider the angular part of the scalar field fluctuations in TNAdS$_4$ in Section \ref{sec:angsec}. In particular, we study in detail how solutions of the angular equations are organized into representations of the isometry algebra, and consider separately the case of a visible/invisible Misner string.
In Section \ref{sec:radialsec} we consider the radial part of the solutions, and show analytically that in the case of a physical Misner string, quasi-normal modes of the scalars will have complex frequencies in the lower half complex plane and thus are stable. Section \ref{sec:discuss} contains a discussion and the outlook of our results. The Appendices contains further technical details.

\section{Kinematics of the  fluid at the boundary of Taub-NUT AdS$_4$}\label{sec:kinematics}

\subsection{General analysis}
An analysis of the holographic fluid at the boundary of TNAdS$_4$ with spherical horizon was presented for the first time in \cite{Leigh:2011au,Leigh:2012jv}. It was  shown  there that the boundary system can be  identified with a perfect  fluid rotating with constant vorticity. We review here its salient  properties and contrast it with the holographic rotating fluid at the boundary of KAdS$_4$.  

The Lorentzian TNAdS$_4$ metric is
\be
\label{TNAdS4}
ds^2=\frac{dr^2}{V(r)}+(r^2+n^2)d\Omega_2^2-V(r)[dt+2n(1-\cos\theta)d\phi]^2\,,
\ee
where $d\Omega_2^2=d\theta^2+\sin^2\theta d\phi^2$ is the usual metric on the unit radius $S^2$ and
\be
\label{V}
V(r)=\frac{1}{r^2+n^2}\left[r^2-n^2-2Mr+\frac{1}{L^2}\left(r^4+6n^2r^2-3n^4\right)\right]\,,
\ee
with $L$ the AdS$_4$ radius. The geometry has an $SU(2)\times \mathbb{R}$ isometry algebra. For generic values of the mass $M>0$ and NUT parameter\footnote{Notice that (\ref{V}) is quadratic in $n$ hence it does not depend on its sign.} $n$ the metric has an outer horizon located at $r_+$ with the topology of a  two-sphere. Its position is given by the largest root of $V(r_+)=0$, namely
\begin{align}
\label{defr+}
r_+[r_+^3+(6n^2+L^2)r_+-2ML^2]=3n^4+n^2L^2\,.
\end{align} 

The holographic analysis of \cite{Leigh:2011au,Leigh:2012jv} yields the conserved, symmetric and traceless boundary energy momentum tensor in the form of a perfect conformal fluid 
\begin{align}
\label{bT}
&T_{\m\n}=p[3u_\m u_\n+g_{\m\n}]\,,\,\,\,p=\frac{M}{8\pi G_4 L^2}\,,\,\,\,\m,\n=0,1,2\,,\\
&\label{Tconserv}
\nabla^\m T_{\m\n}=g^{\m\n}T_{\m\n}=0\,,\,\,T_{\m\n}=T_{\n\m}\,.
\end{align}
with $G_4$ the four-dimensional Newton's constant. The standard holographic interpretation is that (\ref{bT}) corresponds to the expectation value of the energy momentum tensor in the boundary fluid state.
However, this is not the only piece of information that we have regarding the boundary system. In the case at hand, the boundary metric  $g_{\m\n}$ is a particular case of a Papapetrou-Randers (PR) stationary metric\footnote{The generic three-dimensional stationary metric can be written in the PR form $ds^2=-[\Omega dt-b_idx^i]^2+a_{ij}dx^idx^j$, with $i,j=1,2$, $x^\mu=(t,\theta,\phi)$ and $\Omega,b_i,a_{ij}$ functions of $\theta$ and $\phi$.}
\be
\label{bg}
ds^2_{bdy}=-[dt+2n(1-\cos\theta)d\phi]^2+L^2d\Omega_2^2\,.
\ee
In contrast to the vast majority of the examples considered in the fluid/gravity literature the boundary metric (\ref{bg})  is {\it not} conformally flat, having a non-zero Cotton tensor given by
\be
\label{TNCotton}
C_{\m\n}=\frac{n}{L^4}\left(1+\frac{4n^2}{L^2}\right)[3u_\m u_\n+g_{\m\n}]\,.
\ee
Notice that (\ref{TNCotton}) is also of a perfect fluid form which is the reason why TNAdS$_4$ was classified as a perfect geometry in \cite{Mukhopadhyay:2013gja}.

Having (\ref{bg}) as boundary metric results in non-trivial kinematics of the boundary fluid. The latter is determined by its flow velocity $\check{u}=\partial_t$ which is a geodesic, shearless and expansionless congruence of the boundary metric (\ref{bg}) with nonzero vorticity. Explicitly we have\footnote{Recall the definitions of acceleration $\a_\m=u^\n\nabla_\n u_\m$, expansion $\Theta=\nabla_\m u^\m$, shear $\sigma_{\m\n}=h_\m^{\,\,\sigma}h_{\n}^{\,\,\rho}(\nabla_\sigma u_\rho+\nabla_\rho u_\sigma)/2-h_{\m\n}h^{\sigma\rho}(\nabla_\sigma u_\rho)/2$ and vorticity $\omega_{\m\n}=h_\m^{\,\,\sigma}h_{\n}^{\,\,\rho}(\nabla_\sigma u_\rho-\nabla_\rho u_\sigma)/2$, where $h_{\m\n}=g_{\m\n}+u_\m u_\n$. One can interpret $u$ as a gauge field, $\omega$ as the corresponding field strength, and ${\cal C}$ (eq. \eqref{circulation}) as a charge.}
\begin{align}
\label{velocity}
&u^\m=(1,0,0)\,,\,\,u_\m=(-1,0,-2n(1-\cos\theta))\,,\\
\label{vorticity}
&u^\n\nabla_\n u_\m=\nabla_\m u^\m=\sigma_{\m\n}=0\,,\,\,\omega^{TN}_{\m\n}=\left(\begin{array}{ccc}0&0&0\\0&0&-n\sin\theta\\0&n\sin\theta&0\end{array}\right)\,.
\end{align}
In this description the boundary fluid is {\it comoving}  on the stationary PR metric.  A similar description is possible for the holographic fluid at the boundary of the general Kerr-Taub-NUT AdS$_4$ metric. Moreover, in \cite{Leigh:2011au,Leigh:2012jv} it was shown that it is possible to define a natural {\it rotating} frame for the holographic fluids above. This is the Zermelo frame which can be viewed as the frame where the velocity one-form becomes $\hat{u}=dt/\sqrt{\gamma}$ where $\gamma=1-\vec{v}\cdot\vec{v}$ is a Lorentz factor depending on the relative spatial velocity $\vec{v}$ between the PR and Zermelo frames. While the Zermelo frame is everywhere well-defined for KAdS$_4$, it becomes singular above a certain value of the $\theta$ angle for TNAdS$_4$.

\subsection{The total vorticity flux at the boundary}

Vorticity plays an important role in the description of non-relativistic fluids as it enters Kelvin's circulation theorem. The latter states that the circulation, which is equivalent to the vorticity flux through any open surface of an inviscid and barotropic fluid, is constant along the flow. This is similar to the corresponding situation with the magnetic flux in electromagnetism, which implies that in many ways vorticity resembles a magnetic field. The role of vorticity in relativistic fluid dynamics has been emphasised in particular by the work of Carter and Lichnerowicz (see e.g., \cite{Markakis:2016udr,Rezolla}). In particular, for our holographic TNAdS$_4$ fluid the so-called Carter-Lichnerowicz equation takes the simple form
\be
\label{CL}
u^\m\omega^{TN}_{\m\n}=0\,.
\ee
Moreover, by Stoke's theorem the relativistic generalization of the fluid's circulation along a closed path $\gamma$ is given by the vorticity flux through the open surface ${\cal S}$, bounded by $\gamma$ as 
\be
\label{circulation}
{\cal C}=\oint_{\gamma} dx^\m\ u_\m=2\iint_{{\cal S}}dS^{\m\n}\,\omega_{\m\n},
\ee
where $dS^{\m\n}$ is the surface element. For inviscid, barotropic fluids the circulation is constant and this is easily verified along any closed path for the TNAdS$_4$ fluid. 

Since vorticity is a measure of rotation, we see that the holographic TNAdS$_4$ fluid is a rotating fluid. However, it is a very special kind of rotating fluid as can be seen by contrasting it with another known rotating holographic fluid, the one at the boundary of Kerr-AdS$_4$.\footnote{For various properties of the KAdS$_4$ metric see e.g., \cite{Leigh:2011au,Leigh:2012jv}.} The energy momentum of the latter is also of the perfect fluid form (\ref{bT}), and it lives on the three-dimensional Papapetrou-Randers-like metric 
\be
\label{PRKerr}
ds^2=-\left[dt+\frac{a}{\Xi}\sin^2\theta d\phi\right]^2+a_{ij}dx^idx^j\,,\,\,\,a_{ij}=L^2{\rm diag}\left(\frac{1}{\Delta_\theta},\frac{\Delta_\theta}{\Xi^2}\sin^2\theta\right)\,,
\ee
where $|a|\leq L$ is the rotation parameter, $\Delta_\theta=1=a^2\cos^2\theta/L^2$ and $\Xi=1-a^2/L^2$. In contrast to (\ref{bg}) this is a conformally flat metric hence its Cotton tensor vanishes. The fluid's velocity is now 
\be
\label{velocityKerr}
u^\m=(1,0,0)\,,\,\,\,u_{\m}=\left(-1,0,\frac{a}{\Xi}\sin^2\theta\right)\,.
\ee
This is also geodesic, shearless and expansionless, while it has vorticity given by
\be
\label{vorticityKerr}
\omega^{K}_{\m\n}=\left(\begin{array}{ccc}0&0&0\\0&0&\frac{a}{2\Xi}\sin2\theta\\0&-\frac{a}{2\Xi}\sin2\theta&0\end{array}\right)\,.
\ee
The Carter-Lichnerowicz equation (\ref{CL}) is clearly satisfied and the fluid's circulation is constant along any closed path.  

Even though the TNAdS$_4$ and the KAdS$_4$ fluids are both locally rotating they have very different {\it global} rotation properties. The latter can  be studied if one considers their corresponding total vorticity flow. Notice that such a quantity makes sense in the present case where the relevant fluid resides on a two-dimensional spatial surface which is also a compact manifold of finite area. By the usual Stoke's theorem this is generically zero, as in the case of the KAdS$_4$ fluid 
\be
\label{totalfluxK}
{\cal C}^{K}_{tot}=2\oiint_{{\cal S}}dS^{\m\n} \omega^K_{\m\n}=\frac{a}{\Xi}\int_{0}^{2\pi}\!\!d\phi\int_0^\pi d\theta\,\sin 2\theta =0\,.
\ee
The physical meaning of this that when we consider the KAdS$_4$ fluid as a whole the possible sources and sinks of vorticity compensate each other in the sense of rotation i.e., one rotates clockwise and the other counterclockwise much like the Earth's athmosphere appears to be rotating due to the Coriolis effect. We might then say that the KAdS$_4$  fluid {\it does not rotate as a whole}. On the other hand the total circulation of the TNAdS$_4$ fluid is
\be
\label{totalfluxTN}
{\cal C}^{TN}_{tot}=2\oiint_{{\cal S}}dS^{\m\n}\omega^{TN}_{\m\n}=-2n\int_{0}^{2\pi}\!\!d\phi\int_0^\pi d\theta\,\sin \theta =-8\pi n\,.
\ee
The nonzero result is due to the singularity of the TN velocity (\ref{velocity}) and hence of vorticity itself (\ref{vorticity}) at $\theta=\pi$. The singularity may be avoided (see e.g. \cite{Balian:2005joa}) by using at least two different nonsingular velocity fields, related by a total derivative (i.e., a gauge transformation),  in order to describe the flow over the whole boundary spatial surface; in that interpretation, \eqref{velocity} is valid in one such coordinate patch that does not contain $\theta=\pi$. This is of course the exact analogue of the usual magnetic monopole situation, with $-2n$ playing the role of the magnetic charge. We conclude that there is always a non-trivial  source (or sink if we were to take $n<0$) of vorticity over the compact total surface of the TNAdS$_4$ fluid. This source can be conveniently pushed either to spatial infinity or to the origin if we zoom correspondingly to the north or the south pole of the boundary geometry, (\ref{bg}) where we find the Som-Raychaudhuri metric \cite{Som:1968aa}. In particular, for $\theta\mapsto 0,\pi$ the fluid one-form velocity becomes
\be
\label{velocity_phi}
\hat{u} \xrightarrow{\theta\to 0} -dt+\theta n L\hat{e}^\phi+O(\theta^3)\,,\,\,\,\hat{u} \xrightarrow{\theta\to \pi} -dt+\frac{nL}{\pi -\theta}\hat{e}^\phi+O(\pi-\theta)\,,\,\,\,\,\,\,\hat{e}^\phi=\sin\theta d\phi\,,
\ee
and describes rigid rotation near the north pole and an irrotational "bathtub" vortex near the south pole. This can be contrasted with the corresponding behaviour of the KAdS$_4$ fluid 
\be
\label{Kerr_velocity_phi}
\hat{u}^K \xrightarrow{\theta\to 0} -dt+\theta\frac{a}{\Xi}\hat{e}^\phi+O(\theta^3)\,,\,\,\,\hat{u}^K \xrightarrow{\theta\to \pi} -dt+(\pi-\theta)\frac{a}{\Xi}\hat{e}^\phi+O((\pi-\theta)^3)\,,
\ee
which describes rigid rotation in both the north and the south poles. 

\subsection{Making sense of the TNAdS$_4$ fluid}

Our discussion of the velocity (\ref{velocity}) and the vorticity (\ref{vorticity}) of the TNAdS$_4$ fluid is the analogue of Dirac's treatment of the monopole gauge potential and magnetic field \cite{Balian:2005joa} where a singular gauge potential gives rise to a regular magnetic field. However, fluid properties such as vorticity are potentially measurable quantities and hence a careful consideration of their singularities is required. 

To this end, being unaware of analogous discussions in relativistic hydrodynamics, we can turn for guidance to standard nonrelativistic fluid dynamics e.g. \cite{MR1872661} where the standard assumption is that flows on compact manifolds with regular velocity and vorticity, such as a two-sphere, are subject to the {\it Gauss constraint} that sets to zero the total vorticity flux.\footnote{In other words, the vorticity two-form is taken to be globally exact.} In such a case the single point vortex on the two-sphere is singular and to obtain a steady flow on a compact surface one needs to consider two or more point vortices.\footnote{The dynamics of such systems are studied in the context of the so-called $N$-vortex problems \cite{MR1831715}. One may conjecture that the KAdS$_4$ holographic fluid corresponds to the two-vortex system. It would be interesting to examine whether there are gravitational solutions giving rise to known stable $N$-vortex configurations.} On the other hand, single vortices appear regularly in superfluid flows \cite{sonin_2016} where their stability is guaranteed by topological considerations. With this in mind we suggest that there are two complementary ways to make sense of the boundary TNAdS$_4$ fluid:
\begin{itemize}
\item If we require that the fluid lives on a {\it compact} spatial surface at the boundary then we may tolerate having a singular velocity field such as (\ref{velocity}) since the homogeneity of the TNAdS$_4$ spacetime can be used to argue that there is no physical meaning to the boundary point where the velocity diverges. In the magnetic monopole case this is equivalent to stating that the gauge potential does not have physical implications. Nevertheless, the vorticity (\ref{vorticity}) and the total vorticity flux are both globally well defined. Thus, we are describing a system that carries a nonzero total {\it vortex charge} and resides on a compact spatial manifold. This is the case where the bulk {\it Misner string is invisible},\footnote{Note that the form of the metric, or of the velocity form, suggests that we should interpret it as valid on the coordinate patch $\theta<\pi$. When we say the Misner string is invisible, we mean that we are using a particular singular gauge. Perhaps better would be to introduce a second description valid on the patch $\theta>0$, with the two descriptions related on their overlap by a non-trivial transition function. If such a smooth bundle can be constructed over the sphere, we say that the Misner string is invisible. If this is not the case, then the position of the Misner string(s) is physical. Throughout this paper, when we consider a physical Misner string, we take this to mean that there is a single Misner string and we interpret the metric to mean that we have chosen coordinates such that the string is at the South pole, $\theta=\pi$.} which requires a quantization condition. In this case  it is natural to suggest that the boundary system describes a superfluid.
\item On the other hand, we may choose to excise from the boundary the point where the velocity field (\ref{velocity}) diverges, e.g., the south pole. Then our fluid lives on a non-compact spatial manifold with area ${\cal S}$ and has constant vorticity everywhere except at the south pole. We can now add the missing point at the boundary manifold, assign to it a singular vorticity which is such that the total vorticity on the compact space vanishes, namely we take the vorticity two-form to be\footnote{We denote by $\delta_2(\theta-\pi)$ the two-form that has support only at $\theta=\pi$ and $\oiint_{\cal S}\delta_2(\theta-\pi)={\cal S}_{total}$.}
\be
\label{SingularOmega}
{\bf \tilde{\omega}}^{TN}=-2n d\theta\wedge sin\theta d\phi+2n\delta_2(\theta-\pi)\,\Rightarrow\, \oiint_{{}\cal S}{\bf \tilde{\omega}}=0\,.
\ee
This way we can satisfy the Gauss constraint on ${\cal S}$ and we can expect that our boundary system behaves as an ordinary dissipative fluid. In the monopole picture this is equivalent to considering the Dirac string as a physical object. Analogously, we say that this point of view corresponds to taking the {\it Misner string to be physical}. 
\end{itemize}

In the next section, we will investigate how scalar field fluctuations may be constructed in each of these two cases. 

\section{Scalar  field fluctuations in  TNAdS$_4$}
\label{sec:scalars}

\subsection{Setup}
\label{sec:scalarsetup}

To proceed with the analysis of the boundary fluid one needs to study fluctuations and in this paper we consider the simplest case of scalar field fluctuations. By the standard AdS/CFT dictionary \cite{Son:2002sd} the dissipative properties of the holographic fluid can be read from the quasinormal modes of the fluctuating bulk fields, scalars, gauge fields or the metric itself, as the former correspond to the poles of the corresponding retarded Green functions in the boundary.  The equation of motion of a real scalar field $\Phi$ with mass $m_\Phi$ propagating in the TN-AdS$_4$ background is 
\be
\label{Phieom}
\frac{1}{\sqrt{-g}}\partial_\mu[\sqrt{-g}g^{\mu\nu}\partial_\nu\Phi]-m_\Phi^2\Phi=0\,.
\ee
Scalar fluctuations of the metric satisfy this equation for $m_\Phi=0$, but we will keep this parameter for the present.
The TNAdS$_4$ geometry as given in eq. \eqref{TNAdS4} (a more detailed discussion of TNAdS$_4$ spacetimes is presented in Appendix A) with spherical horizon has an $SU(2)\times \mathbb{R}$ isometry generated by the vector fields
\beqn
\label{TNAdS4Kill1}
{\boldsymbol \xi}_1 &=& -\sin\phi \cot\theta\,\partial_\phi +\cos\phi\,\partial_\theta -2n\sin\phi\frac{1-\cos\theta}{\sin\theta}\,\partial_t\,, \\
{\boldsymbol \xi}_2 &=& \cos\phi \cot\theta\,\partial_\phi + \sin\phi\,\partial_\theta +2n\cos\phi\frac{1-\cos\theta}{\sin\theta}\,\partial_t\,, \\
{\boldsymbol \xi}_3 &=& \partial_\phi -2n\partial_t,\qquad\qquad {\bf e} = \partial_t
\eeqn
We notice that non-zero NUT charge has led to a twisting of the generators ${\boldsymbol \xi}_{1,2}$ by a vector field proportional to $\pa_t$. Since the metric is $\phi$- and $t$-independent, any linear combination of $\pa_\phi$ and $\pa_t$ with constant coefficients is an isometry, and we have chosen ${\boldsymbol \xi}_3$ as above such that the Lie brackets are diagonalized, viz,
\beq
[{\boldsymbol \xi}_i,{ \boldsymbol \xi}_j ] =-\epsilon_{ijk}{ \boldsymbol\xi}_k\qquad[{\boldsymbol \xi}_i,{\bf e}]=0\quad i,j,k=1,2,3. 
\eeq
That is, ${\boldsymbol \xi}_3$ is an $SU(2)$ generator.
One of the important aspects of the isometry algebra is that the $SU(2)$ orbits are not closed, but take a helical form whose pitch is proportional to the NUT charge.
We will write solutions to the scalar equations of motion by diagonalizing ${\boldsymbol \xi}_3$ and ${\bf e}$. 
Writing  $\Phi(t,r,\theta,\phi)=e^{-i\omega t}f(r,\theta,\phi)$ we obtain after some rearrangement 
\be
\label{feom}
\left\{\partial_r[(r^2+n^2)V(r)\partial_r]-(r^2+n^2)\left[m_\Phi^2-\frac{\omega^2}{V(r)}\right]+4n^2\omega^2-{\bf L}^2\right\} f(r,\theta,\phi)=0\,,
\ee
with
\be
\label{L2}
{\bf L}^2 =-\frac{1}{\sin^2\theta}\left[\sin\theta\,\partial_\theta(\sin\theta\partial_\theta)+\left(\partial_\phi+i2n\omega(1-\cos\theta)\right)^2\right]+4n^2\omega^2\,.
\ee
Under the identifications $e\equiv \omega$ and $ g\equiv -2n$, ${\bf {L}}^2 $  coincides with the square of the generalized angular momentum operator\footnote{The generalized angular momentum operator $\vec{L}$ of a particle with charge $e$ in a monopole background with gauge potential (we take the north-hemisphere representative here) $\vec{A}=\frac{g}{r}\frac{1-\cos\theta}{\sin\theta}{\bf \hat{e}_\phi}$, is given by $\vec{L}=\vec{L}_{c}-eg{\bf \hat{r}}$, where the canonical momentum is  $\vec{L}_{c}=\vec{r}\times (-i\hbar\vec{\nabla}-e\vec{A})$ and ${\bf \hat{e}_\phi}$, ${\bf \hat{r}}$ are unit vectors. The generalized angular momentum satisfies the usual $SU(2)$ commutation relations. }  for a particle of charge $e$ in the background of a monopole of charge $g$  \cite{Balian:2005joa}. 
Finally, we separate variables as $f(r,\theta,\phi)=R(r)Y(\theta,\phi)$ to obtain the set of equations 
\begin{align}
\label{radialEq1}
&\left\{\frac{d}{dr}[(r^2+n^2)V(r)\frac{d}{dr}]-(r^2+n^2)\left[m_\Phi^2-\frac{\omega^2}{V(r)}\right]+4n^2\omega^2-C\right\} R(r)=0\,,\\
\label{angularEq1}
&\left\{ {\bf L}^2\,-C \right\}Y(\theta,\phi)=0\,.
\end{align}
As written, the separation constant $C$ will play the role of the quadratic Casimir of $SU(2)$. 

\subsection{The angular equation and $SU(2)$ modules}
\label{sec:angsec}

In the context of holographic hydrodynamics one is mainly interested in the radial equation (\ref{radialEq1}) whose spectrum would unveil the physics of the boundary system, given prescribed boundary conditions. However, in the study of TNAdS$_4$ we are forced to deviate from this procedure as we need to discuss in detail the angular equation (\ref{angularEq1}) first. In other words we need to settle the issue of the holographic interpretation of the Misner string singularity of TNAdS$_4$. 

In the absence of a cosmological constant it was argued by Misner \cite{Misner:1963fr} that the string singularity of Taub-NUT spacetimes can be made invisible if the time coordinate is compactified. Although perhaps not a problem in the Euclidean continuation, in Lorentzian signature this implies that Taub-NUT spacetimes have closed timelike curves (CTCs) which in turn raises serious questions regarding their possible physical relevance. To avoid this pathology one therefore has to take the approach that the Misner string is a physical object of the bulk spacetime and hence it produces physical effects in the holographic boundary. We will consider below both points of view, namely  a physical and an invisible Misner string in the bulk of TNAdS$_4$, and discuss their distinct consequences for the boundary fluid. 
Our calculation extends and completes the one presented in \cite{Leigh:2011au}.

It is convenient to write the eigenvalue of the quadratic Casimir as $C=q(q+1)$. We will seek a basis of solutions of the angular equation such that ${\bf L}_3=-i{\boldsymbol \xi}_3$ acts diagonally, with eigenvalue $m$. As such, we have
\beq
\label{PhiSep}
Y(\theta,\phi)\equiv Y_{q,m,\Omega}(\theta,\phi)=Y_{q,m,\Omega}(\theta)e^{i(m-\Omega)\phi},
\eeq
where we have set $\Omega=2n\omega$. We note the all-important feature of such solutions, that {\it the frequency appears in the azymuthal dependence} given the group theory interpretation of $m$, and this will have an important effect on the nature of the solutions. This is related to the aforementioned fact that the $SU(2)$ orbits are not closed in general and it is in sharp contrast with the corresponding calculation in the Kerr AdS$_4$ case \cite{Uchikata:2009zz,Cardoso:2013pza} where the azimuthal dependence is of the usual form $e^{im\phi}$ and allows  for the common assumption of real $m$ and complex $\omega$.    We note that the solutions satisfy 
\beq
\label{periodicity}
\Phi(t,r,\theta,\phi+2\pi)=e^{2\pi i(m-\Omega)}\Phi(t,r,\theta,\phi).
\eeq
Clearly the fate of periodicity properties of the solutions rests on the value of $\Omega$ (which ultimately will come from solving the radial equation) and on the values of $m$.
Given the form of eq. \eqref{PhiSep}, eq. \eqref{angularEq1} then becomes
\beq
\label{angularEq2}
\left\{\frac{1}{\sin\theta}\frac{d}{d\theta}\left[\sin\theta\frac{d}{d\theta}\right]-\frac{(m-\Omega\cos\theta)^2}{\sin^2\theta}-\Omega^2+q(q+1)\right\}Y_{q,m,\Omega}(\theta)=0\,.
\eeq

We begin by considering the invisibility of the Misner string singularity. 
As was mentioned above,  eq. \eqref{angularEq2} coincides with eq. (22) of Wu and Yang \cite{Wu:1976ge} if we make the identifications $e\equiv \omega$ and $ g\equiv -2n$. Thus we have that the NUT charge plays the role of magnetic charge, and the frequency the electric charge.  If we then simply repeat their analysis, we would conclude that the $Y_{q,m,\Omega}(\theta,\phi)$ give a complete, everywhere regular and normalizable set of eigenfunctions of ${\bf {L}}^2$ in (\ref{angularEq1}), which are the so-called monopole harmonics, a generalization of the more familiar spherical harmonics \cite{Wu:1976ge,Balian:2005joa}.

However, we should emphasize an important difference of our analysis with respect to all the previous  cases that we are aware of, where equations (\ref{angularEq1}) and (\ref{angularEq2}) have appeared in the past. In all those cases the parameter $\Omega$ was identified with the product of the electric and the monopole charge and hence it was {\it a priori assumed to be quantized \`a  la Dirac}, namely $\Omega\in \mathbb{Z}$, or equivalently $\omega=2\pi\frac{k}{4\pi n}, k\in\ZZ$. Such a quantization argument was based on quantum mechanical considerations. In contrast, here $\Omega$ is the  parameter that determines the nature of the excitations of the boundary fluid, and as such it needs to be evaluated using the physical conditions that we will impose on our system, namely we would need to solve the radial equation (\ref{radialEq1}) with the appropriate boundary conditions, and it is well-known that, at least in the presence of dissipation, {\it the allowed frequencies are complex}. Clearly, this is in stark contrast to $\Omega$ being an integer! We conclude that the invisibility of the Misner string could only be possible in the absence of dissipation. What we will show below is that in the presence of dissipation, a scalar field with complex frequency may be organized into representations of the isometry algebra, but these representations are not the familiar unitary finite dimensional representations.

To proceed, it is convenient to introduce the coordinate
\beq
u=\sin^2(\theta/2)\,.
\eeq
The domain $\theta<\pi$ is thus mapped to the open disk $u<1$. Then if we introduce
\beq
{\cal N}\equiv m-\Omega,
\eeq
eq. (\ref{angularEq2}) becomes
\beq
\label{angularEq3}
\left\{\frac{d}{du}\left[u(1-u)\frac{d}{du}\right]+\left[q(q+1)-\Omega^2-\frac{({\cal N}+2u\Omega)^2}{4u(1- u)}\right]\right\}Y_{q,m,\Omega}(u)=0\,.
\eeq
For arbitrary values of $q,m,\Omega$, this can be cast as a hypergeometric equation, and the solutions can be written in the form
\beq\label{genSolPhi}
Y_{q,m,\Omega}(u)=u^{a/2}(1-u)^{b/2}{}_2F_1(1+q+\tfrac{a+b}{2},-q+\tfrac{a+b}{2};1+a;u)\,,
\eeq
where $a=\pm{\cal N}=\pm(m-\Omega), b=\pm(2m-{\cal N})=\pm(m+\Omega)$. Because of functional equalities satisfied by hypergeometrics, these four functions are not all independent, and for our purposes, it is sufficient to take
\beq
Y^\pm_{q,m,\Omega}(u)=u^{\pm{\cal N}/2}(1-u)^{\pm(2m-{\cal N})/2}{}_2F_1(1+q\pm m,-q\pm m;1\pm {\cal N};u)\,.
\eeq
 To summarize we have cast the solutions to the scalar wave equation in the form
\beqn
\label{sepvar}
\Phi^\pm_{q,m,\Omega}(t,r,u,\phi)&=&R_{q,\Omega}(r)\Psi^\pm_{q,m,\Omega}(t,u,\phi)\,,\\
\Psi^\pm_{q,m,\Omega}(t,u,\phi)&=&e^{-i\omega t}Y^\pm_{q,m,\Omega}(u,\phi)=e^{-i\omega t}e^{i{\cal N}\phi}Y^\pm_{q,m,\Omega}(u)\,.
\eeqn
As usual, it is convenient to write the other two $SU(2)$ generators in complexified form as ${\bf L}_\pm =\pm{\boldsymbol \xi}_1+ i {\boldsymbol\xi}_2 $, whereby
\beq
{\bf L}_\pm = \frac{ie^{\pm i\phi}}{\sqrt{u(1- u)}}\left[ \mp iu(1- u)\partial_u +\frac{1-2 u}{2}\partial_\phi+2nu\partial_t\right] \,.
\eeq
Moreover, we obtain
\beqn
\label{su2pluslower}
 {\bf L}_-(\Psi^+_{q,m,\Omega}(t,u,\phi))&=&-{\cal N}\Psi^+_{q,m-1,\Omega}(t,u,\phi)\,,\\
\label{su2up}
{\bf L}_+(\Psi^+_{q,m,\Omega}(t,u,\phi))&=&\frac{(1+q+m)(m-q)}{1+{\cal N}}\Psi^+_{q,m+1,\Omega}(t,u,\phi)\,,\\
\label{su2down}
{\bf L}_-(\Psi^-_{q,m,\Omega}(t,u,\phi))&=&\frac{(1+q-m)(m+q)}{1-{\cal N}}\Psi^-_{q,m-1,\Omega}(t,u,\phi\,,)\\
\label{su2minusraise}
 {\bf L}_+(\Psi^-_{q,m,\Omega}(t,u,\phi))&=&-{\cal N}\Psi^-_{q,m+1,\Omega}(t,u,\phi)\,.
\eeqn
From these, we can also deduce that
\beqn
 {\bf L}_+{\bf L}_-(\Psi^\pm_{q,m,\Omega}(t,u,\phi))&=&\Big(q(q+1)-m(m-1)\Big)\Psi^\pm_{q,m,\Omega}(t,u,\phi)\,,\\
{\bf L}_- {\bf L}_+(\Psi^\pm_{q,m,\Omega}(t,u,\phi))&=&\Big(q(q+1)-m(m+1)\Big)\Psi^\pm_{q,m,\Omega}(t,u,\phi)\,.
\eeqn
From the above we finally obtain
\beqn
&i{\bf e}(\Psi^\pm_{q,m,\Omega}(t,u,\phi))=\omega\Psi^\pm_{q,m,\Omega}(t,u,\phi)\,,\\
&{\bf L}_3(\Psi^\pm_{q,m,\Omega}(t,u,\phi))=m\Psi^\pm_{q,m,\Omega}(t,u,\phi)\,,\\
& {\bf L}^2(\Psi^\pm_{q,m,\Omega}(t,u,\phi))=q(q+1)\Psi^\pm_{q,m,\Omega}(t,u,\phi)\,.
\eeqn
Thus, we see that the functions $\Psi^\pm_{q,m,\Omega}(t,u,\phi)$ will in fact give representations of $SU(2)\times\mathbb{R}$. Again, all of these results have been established without specifying what values $q,m,\Omega$ might take. 

As mentioned previously, the periodicity properties of the solutions in $\phi$ are determined by the values of ${\cal N}=m-\Omega$, and we furthermore now see that the behaviour of the solutions at $u\sim 0$ and $u\sim 1$ are also determined by $m,\Omega$.\footnote{Notice that $m-\Omega$ is the eigenvalue of $-i\pa_\phi$, while $m+\Omega$ is the eigenvalue of $-i(\pa_\phi-4n\pa_t)$. The metric norm of these two vector fields vanishes at $u=0,1$ respectively. The form of the solution given above, and in particular its dependence on $m,\Omega$, follows from having taken the metric in a form which places the single Misner string at the South Pole, $u=1$.} Clearly, if we were to require that the solutions be single-valued in $\phi$, then we would require ${\cal N}=m-\Omega\in\mathbb{Z}$, which we write as 
\beq
m=\Omega+k,\qquad k\in\mathbb{Z}.
\eeq
When ${\cal N}>0$, for the solutions to be finite at $u=0$, we must take only the solutions  $\Psi^+_{q,m,\Omega}(t,u,\phi)$.\footnote{Similarly, if ${\cal N}<0$, it is $\Psi^-_{q,m,\Omega}(t,u,\phi)$ that are finite at the origin.} Then, in order for the solutions to be finite at $u=1$, we must also have $m+\Omega>0$, which implies that
\[ m=\alpha+k,\qquad \Omega=\alpha,\qquad 2\alpha\in\mathbb{Z}\,, \]
for some integer $k$.
These requirements lead to the monopole harmonics, and in the special case $\alpha=0$ to the spherical harmonics. In this case, we see that $\Omega$ must be quantized, all of the solutions are well-defined everywhere on $S^2$ and the multiplets are finite dimensional. Indeed, we can see from eqs. \eqref{su2up} that $\Psi^+_{\ell,\ell,\ell-k}(t,u,\phi)$ is the $SU(2)$ highest weight state in the $q=\ell$ multiplet, with $\ell$ being integer or half-integer, while $\Psi^-_{\ell,-\ell,\ell-k}(t,u,\phi)$ is the lowest weight state.

That is a perfectly fine basis of functions as long as the radial equation returns to us real and quantized values of $\omega$. But what if it doesn't? Suppose that the radial equation gives us a complex frequency $\omega$. Given that quasi-normal mode, there will be another mode  (by parity invariance) with frequency of the form $\bar\omega=-\omega^*$. Indeed, note that if $Y_{q,m,\Omega}(u,\phi)$ is a solution of eq. \eqref{angularEq1}, then $(Y_{q,m,\Omega}(u,\phi))^*$ solves the same equation with $(q,m,\Omega)$ replaced by the data $(\bar q,\bar m,\bar\Omega)=(q^*,-m^*,-\Omega^*)$. 
\beqn
Y_{q,m,\Omega}(u,\phi)&=&
e^{i(m-\Omega)\phi} u^{(m-\Omega)/2}(1-u)^{(m+\Omega)/2} {}_2F_1(1+q+m,-q+m,1+m-\Omega;u)\,,
\\
(Y_{q,m,\Omega}(u,\phi))^*&=&
e^{-i(m^*-\Omega^*)\phi} u^{(m^*-\Omega^*)/2}(1-u)^{(m^*+\Omega^*)/2} {}_2F_1(1+q^*+m^*,-q^*+m^*,1+m^*-\Omega^*;u)\,,\nonumber
\eeqn
and we will write the latter as
\beq
\tilde Y_{\bar q,\bar m,\bar\Omega}(u,\phi)=e^{i(\bar m-\bar\Omega)\phi} u^{-(\bar m-\bar\Omega)/2}(1-u)^{-(\bar m+\bar\Omega)/2} {}_2F_1(1+\bar q-\bar m,-\bar q-\bar m,1-\bar m+\bar\Omega;u)\,.
\eeq
Hence we have
\beq
{\bf  L}_-\left(e^{-i\bar{\omega}t}\tilde Y_{\bar q,\bar m,\bar\Omega}(u,\phi)\right)=\frac{(1+\bar q-\bar m)(-\bar q-\bar m)}{1-\bar m+\bar\Omega}e^{-i\bar{\omega}t}\tilde Y_{\bar q,\bar m-1,\bar\Omega}(u,\phi)\,,
\eeq
and
\beq
{\bf L}_+\left(e^{-i\omega t}Y_{q,m,\Omega}(u,\phi)\right)=\frac{(1+q+m)(m-q)}{1+m-\Omega}e^{-i\omega t}Y_{q,m+1,\Omega}(u,\phi)\,.
\eeq
 We thus see that $Y_{q,q,\Omega}(u,\phi)$ gives rise to the highest weight state (hws) $\Psi_{q,q,\Omega}(t,u,\phi)$ with ${\bf L}_+\Psi_{q,q,\Omega}(t,u,\phi)=0$, while $\tilde Y_{\bar q,-\bar q,\bar\Omega}(u,\phi)$ gives rise to a corresponding lowest weight state (lws) $\tilde{\Psi}_{\bar{q},-\bar{q},\bar{\Omega}}(t,u,\phi)$. Given eqs. \eqref{su2pluslower}--\eqref{su2minusraise}, the $Y_{q,m,\Omega}(u,\phi)$ generally correspond to elements of a highest weight representation (hwr) of $SU(2)$, while $\tilde Y_{\bar q,\bar m,\bar\Omega}(u,\phi)$ are elements of a lowest weight representation (lwr). Thus the quasi-normal modes with frequency $\omega$ are associated with an hwr, while the dual frequency $-\omega^*$ is associated with an lwr. These representations are generally non-unitary, infinite dimensional and irreducible, and indeed the hwr and lwr are not necessarily the same representation (unless finite dimensional) --- the more familiar finite-dimensional representations of $SU(2)$ are in this language both highest- and lowest-weight, because they satisfy a 'quantization' condition that allows the hws and lws to occupy the same multiplet. Indeed upon the imposition of a quantization condition (see eqs. (\ref{su2pluslower}--\ref{su2minusraise})), the two representations can merge into a finite-dimensional self-dual representation. We will provide a consistent picture below in which this does not happen -- complex frequencies lead to infinite dimensional non-unitary representations of the $SU(2)$ algebra.\footnote{The reader will note that the structure of these (non-unitary) $SU(2)$ multiplets are very closely related to multiplets of $SL(2,\mathbb{R})$. This is expected since the two algebras have the same complexification. We do not require that the representations of the algebra that we consider extend to representations of the group $SU(2)$. See \cite{Kitaev:2017hnr} for a recent review of $SL(2,\mathbb{R})$ representations in the context of JT gravity and SYK. We expect that the analogous $SL(2,\mathbb{R})$ multiplets would arise in the context of the AdSTN black hole with hyperbolic horizon.}  
 
Before proceeding further, perhaps we should note that the reader may be surprised by these claims. Usually, one takes the finite unitary irreducible $SU(2)$ representations without further thought, as we are taught to do in quantum mechanics. However, we should note that here we are not solving a quantum mechanics problem, and furthermore, although we have complexified the problem by introducing raising and lowering operators to display the $SU(2)$ structure of solutions, the $SU(2)\times\mathbb{R}$ generators are not self-adjoint in general. This is particularly clear if $\omega$ is complex, implying that $i\pa_t$ cannot be interpreted as a Hermitian operator. Furthermore, since $\pa_t$ is intertwined into the $SU(2)$ generators, they cannot be interpreted as having simple Hermitian properties either. For this reason, we should not expect to obtain a unitary representation. 

Nevertheless, we will now introduce requirements that seem to lead to a consistent picture for any quasi-normal modes. First, we will relax the condition on ${\cal N}$ to simply be that ${\cal N}=m-\Omega$ is real rather than integer-valued.  One might interpret this to mean that $-i\pa_\phi$ is self-adjoint.\footnote{This statement is imprecise, as we have not introduced a notion of inner product. We will make further remarks about this later in the paper. } Since the solutions depend on $\phi$ as $e^{i{\cal N}\phi}$, this condition does not necessarily imply that the solutions are single-valued, but at least they will transform by a pure phase under $\phi\to\phi+2\pi$. We will interpret this to mean that the solutions are of an `anyonic' character, and in this sense detect the presence of the Misner string.

Consider the hwr and lwr found above in light of this assumption. It implies that the values of $m$ and thus $q$ are complex, but their imaginary parts are fixed by the imaginary part of $\omega$. We write $\omega=\omega_1+i\omega_2$ and $\Omega=\Omega_1+i\Omega_2$. Then we have 
\beqn\label{hwrnums}
hwr:&\qquad q=q_1+i\Omega_2,\qquad &m=q_1-k+i\Omega_2,\quad k=0,1,...\\
lwr:&\qquad \bar q=q_1-i\Omega_2,\qquad &\bar m=-q_1+k+i\Omega_2,\quad k=0,1,...
\eeqn
The value of $q_1$ has not been determined, but we note that ${\cal N}=q_1-\Omega_1-k$ and $\bar{\cal N}=-q_1+\Omega_1+k=-{\cal N}$. We also note that the Casimir evaluates to
\beqn
hwr:&\qquad C=q(q+1)=q_1(q_1+1)-\Omega_2^2+i(2q_1+1)\Omega_2\,,\\
lwr:&\qquad \bar C=\bar q(\bar q+1)=q_1(q_1+1)-\Omega_2^2-i(2q_1+1)\Omega_2\,.
\eeqn
We see that the values of the Casimir are complex conjugates of each other.  
Thus if we require that the hwr and lwr are dual representations in the sense that they share the same value for the quadratic Casimir, then $C$ must be real and $q_1=-\frac12$. In this case, $C=-\frac14-\Omega_2^2$. 
\beqn
hwr:&\qquad q=-\frac12+i\Omega_2,\qquad &m=-\frac12-k+i\Omega_2,\qquad {\cal N}=2n\varpi-\frac12-k\,,\\
lwr:&\qquad \bar q=-\frac12-i\Omega_2,\qquad &\bar m=\frac12+k+i\Omega_2,\qquad \bar{\cal N}=-2n\varpi+\frac12+k\,.
\eeqn
Here we have written $\omega_1=-\varpi$ and assume that $\varpi>0$. 
For clarity we have plotted these eigenvalues in Figure \ref{fig:states.pdf}. 

\myfig{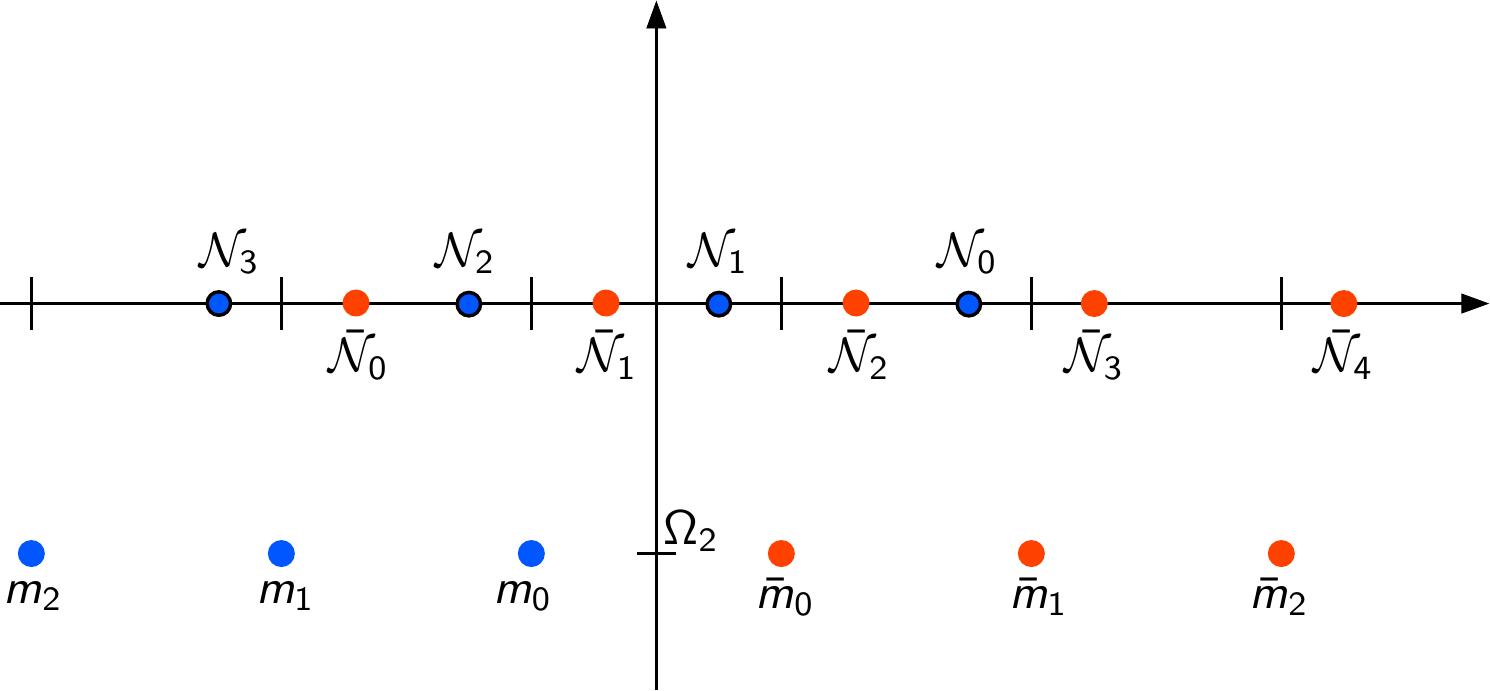}{10}{The hwr and lwr (for real $C$) plotted in the complex plane for a typical value of $\varpi>0$ and $\Omega_2<0$.}

If the hwr and lwr are dual representations, then there is a natural $SU(2)$-invariant inner product (see Appendix).\footnote{Here we are referring to the $SU(2)$-invariant integral of the product of an element of the hwr with an element of the lwr (that is, ${\cal R}\times {\cal R}^*$ contains the identity).} We have assigned the frequencies as above to the hwr such that the hws behaves as $u^{(\varpi-1/2)/2}$ near $u\to 0$. Notice then that for $\varpi>0$, we have integrability at the origin. However, it is inevitable that $SU(2)$ descendants will be singular at the origin. This feature may remind the reader of the Aretakis instability \cite{Aretakis:2012bm,Aretakis:2012ei,Hadar:2017ven,Hadar:2018izi}.  

If we do not require that the hwr and lwr are dual representations, then $q_1$ is free to be any real number, and $C$ is an arbitrary complex number. We will touch upon this further in the next section. We can anticipate though that the radial equation with infalling boundary conditions will yield specific values of $\omega$ which vary continuously with $q_1$. This situation is reminiscent of the analysis that has been done for Kerr-AdS \cite{Uchikata:2009zz}; in that case, there is no $SU(2)$ isometry to organize solutions by, and the separation constant that is the analogue of $C$ was taken to be complex.  We will consider the general case in the next section.
 
\subsection{The radial equation}

In view of the discussion above we now find ourselves in a rather peculiar situation as far as holography is concerned. In a typical AdS/CFT calculation of fluctuations around bulk backgrounds the spectrum of $\omega$ is determined by solving the radial equation and imposing the relevant boundary conditions. For example, in the absence of a horizon in the bulk, one usually obtains just the normal modes e.g., \cite{Balasubramanian:1998sn}, while in the presence of a bulk horizon the physical boundary conditions give rise to the generally complex frequency quasinormal modes, e.g., \cite{Kovtun:2005ev}. In our case, imposing regularity of the solutions (\ref{genSolPhi}) would fix the mode frequencies $\omega$ to be real, hence $\omega^2>0$, and therefore we cannot have quasinormal modes in scalar fluctuations around the TNAdS$_4$ geometry. This is a remarkable result since the imaginary part of quasinormal modes corresponds to dissipation in the boundary theory. This is consistent with our interpretation of the boundary modes as quantum vortices, and strengthens our suggestion that in the absence of a physical Misner string the holographic fluid dual to TNAdS$_4$ is in a superfluid state. Equivalently, one might say that the TNAdS$_4$ with periodic time coordinate is an unusual black hole. 

\subsubsection{The Schr\"odinger Problem}\label{sec:radialsec}

In any case, our suggestions should be consistent with the analysis of the radial equation (\ref{radialEq1}). Let us first consider the problem as a Schr\"odinger system. Setting there 
\beq\label{RtoZ}
R(r)=\frac{1}{\sqrt{r^2 +n^2}}Z(r)\,,
\eeq
we obtain
\be\label{radialEqZ1}
V(r)Z''(r)+V'(r) Z'(r)+\left[\frac{\omega^2}{V(r)}h(r)^2-
 \frac{C}{r^2+n^2}-U_{TN}(r)\right]Z(r)=0\,,
\ee
with
\be 
\label{UTNh}
 U_{TN}(r) =m_\phi^2+
 \frac{rV'(r)}{r^2+n^2}+\frac{n^2V(r)}{(r^2+n^2)^2}\,,\qquad
  h(r)^2=1+\frac{4n^2V(r)}{r^2+n^2}\,,
\ee
where the prime denotes differentiation with respect to $r$. 
 
To bring the radial equation into a Schr\"odinger form,  we first define the tortoise coordinate $r_*$ as
\be
\label{r*}
\frac{dr_*}{dr}=\frac{h(r)}{V(r)}\,\Rightarrow\,r_*\sim \frac{1}{4\pi Tr_+}\ln(r-r_+)+\cdots\,,
\ee
where the ellipsis denotes terms involving positive powers in $(r-r_+)$. To derive (\ref{r*}) we have used (\ref{TNtemp}). In the tortoise coordinate the horizon is at $r_*\rightarrow-\infty$ and the boundary at $r_*\rightarrow\infty$. Finally, we can bring (\ref{radialEqZ1}) into a Schr\"odinger form by introducing $\psi(r)=\sqrt{h(r)}Z(r)$ and we obtain
 \begin{align}
 \label{radialSchr}
 &\frac{d^2}{dr_*^2}\psi(r_*)+\left[\omega^2-{\cal U}_{TN}(r_*)\right]\psi(r_*)=0\,,\\
 \label{VSchr}
 &{\cal U}_{TN}(r_*)=\frac{V(r)}{h(r)^2}\left[U_{TN}(r)+\frac{C}{r^2+n^2}+\frac{1}{2}\frac{V'(r)h'(r)}{h(r)}+\frac{1}{2}\frac{V(r)h''(r)}{h(r)}-\frac{3}{4}\frac{V(r)h'(r)^2}{h(r)^2}\right].
 \end{align}
where in the definition of ${\cal U}_{TN}(r_*)$ we regard $r=r(r_*)$ throughout. As usual in this sort of analysis, $\omega^2$ plays the role of Schr\"odinger energy. If $\omega^2$ were negative, then $\omega$ must be complex, and clearly such a situation is associated with the existence of `bound states' for the Schr\"odinger problem. 
We set $L=1$ and we note that the potential depends on $n,C,r_+$. For generic values of the parameters $r_+$ and $C$,\footnote{For the sake of this discussion, we are taking $C\in\RR$ because it is only under that assumption that we can expect $\omega^2$ to be real. The Schr\"odinger analysis does not then apply to the $C\in\mathbb{C}$ possibility mentioned at the end of the last section. We take the scalar mass above the BF bound, e.g. $m_\phi=0$.} one finds that there is always a critical value $n_*$ of the NUT charge for which the potential vanishes at some distance $r>r_+$. 
Given that $T$ is determined by $r_+$ and $n$, for a given $r_+$, we can associate a critical temperature $T_*$ to the value $n_*$. For $n<n_*$ the potential is always positive outside the horizon and can be thought of as a deformation of the usual Schr\"odinger-like potential of the Schwarzchild AdS$_4$ black hole. This is the low-temperature $T<T_*$ regime which supports the presence of the quantized vortices. For $n>n_*$ we pass to a high-temperature regime where ${\cal U}_{TN}(r_*)$ develops a potential well of finite depth $-U<0$ and width $W>0$. If we approximate the potential with a semi-infinite rectangular well the usual condition for the existence of a bound state is $UW^2\geq \pi^2/8$. This would give a critical temperature $T_{**}>T_*$ above which the system can no longer support quantized vortices, and we are forced to consider the Misner string as physical. 

It should be clear from the analysis above that the Taub-NUT AdS$_4$ fluid is a peculiar case. Indeed, it is not uncommon in AdS/CFT to have a situation where the radial potential for black hole fluctuations becomes negative outside the horizon. The typical example, first discussed by Gubser in \cite{Gubser:2008px}, involves complex scalars in the background of a charged black hole. Is such a case the coupling of the scalars to the background gauge potential results in a negative contribution to the mass $m_{\phi}^2$ of the scalars which may consequently drop below the BF bound, giving rise to an instability of black hole fluctuations in the form of negative energy $\omega^2<0$ bound states.  This is the backbone of holographic superconductivity \cite{ Hartnoll:2008vx, Hartnoll:2008kx}. Our case is very similar to the situation discussed briefly in Appendix A of \cite{Hartnoll:2008kx}, namely that of the instability of a near extremal charged black hole, conformally coupled to a neutral scalar. In the latter case the instability is intimately related to the existence of an AdS$_2\times {\mathbb R}^2$ throat of the extremal charged black hole, such that the mass of the conformal scalar is always below the BF bound of AdS$_4$. Nevertheless, in contrast to this case, we have here an instability which seems to be unrelated to extremality or AdS$_2$ and it is driven by the NUT charge $n$. More intriguingly, $n$ determines the temperature and hence the instability seems to occur for large temperatures.

\subsubsection{Physical Misner string}\label{sec:scalarvis}

Our aim in this section is to briefly discuss the radial equation (\ref{radialEq1}) with infalling boundary conditions at the horizon and Dirichlet boundary conditions at the asymptotic boundary, as is appropriate for the calculation of the spectrum of quasinormal modes. We will leave detailed numerical analysis to a future publication, and content ourselves here with analytical comments. What we will demonstrate is that generically complex frequencies are found when infalling boundary conditions are imposed at the horizon, and so our only conclusion is that the Misner string must be physical in order for dissipation to occur. What we will be most interested in here is whether or not we can ascertain if the system is stable, that is if the quasi-normal mode frequencies are in the lower half complex plane.

We begin with the radial equation \eqref{radialEqZ1}, 
and write
\beq\label{ZPsi}
Z(r)=e^{-i\omega r_*}\Psi(r)\,,
\eeq
where $r_*$ is the tortoise coordinate \eqref{r*}.
We will require that $\Psi(r_+)$ be finite. This brings the radial equation to the form
\beq
V(r)\Psi''(r)+[V'(r)-2i\omega h(r)]\Psi'(r) -\left[i\omega h'(r)+U_{TN}(r)+\frac{C}{r^2+n^2}\right]\Psi(r)=0.
\eeq
Given the redefinitions that we have made, it is natural\footnote{That is, the radial part of any natural inner product would involve that.} to multiply by $\Psi(r)$ and integrate over $r$, obtaining
\beq
\int_{r_+}^{\infty}dr\left\{V(r)\Psi^*(r)\Psi''(r)+[V'(r)-2i\omega h(r)]\Psi^*(r)\Psi'(r)-\left[i\omega h'(r)+U_{TN}(r)+\frac{C}{r^2+n^2}\right]|\Psi(r)|^2\right\}=0\,.
\eeq
A series of standard manipulations involving integration by parts yields
\beq
\int_{r_+}^{\infty}dr\Big\{ V(r)|\Psi'(r)|^2
+V_{TN}(r)|\Psi(r)|^2\Big\}=-\frac{|\omega|^2}{Im\ \omega}|\Psi(r_+)|^2\,,
\eeq
where $V_{TN}(r)=U_{TN}(r)+\frac{Q}{r^2+n^2}$, where $Q=\frac{Im\  C^*\omega}{Im\ \omega}=(q_1-\Omega_1)(q_1-\Omega_1+1)-|\Omega|^2$, using the notation of eq. \eqref{hwrnums}.
We see that if $V(r)$ and $V_{TN}(r)$ were everywhere positive outside the horizon, then the left-hand side is strictly positive and we would conclude that $Im\ \omega < 0$ and thus any quasi-normal mode would be stable. Although $V(r)$ and $U_{TN}(r)$ are positive everywhere, the term involving $Q$ in $V_{TN}(r)$ can be negative. It is simple to see that by plotting $V_{TN}(r)$ for a range of values of $n,r_+,Q$, $V_{TN}(r)$ is in fact positive everywhere. Preliminary numerical analysis indicates that there are indeed stable quasi-normal modes. 
 
\section{Discussion and Outlook}\label{sec:discuss}

In this paper, we have explored scalar field fluctuations in Lorentzian TNAdS$_4$ with spherical horizon. This is a useful playground, because it represents a simple example of an asymptotically {\it locally} AdS geometry which is in an interesting state of the dual boundary theory. Our analysis has been benefitted by the existence of a large isometry algebra. We have found that the physics of the scalar fluctuations depends crucially on the nature of the Misner string singularity. In the case where the Misner string is taken as invisible (analogous to the Dirac strings found in electromagnetism), one arrives at an interpretation of the scalar modes as quantized vortices in a dissipationless fluid. Holographically, such a situation could not apparently support modes falling through the black hole horizon. Given that an invisible Misner string has conceptual problems, we also considered the case in which the Misner string is taken to be a physical object. We have shown that this case does support the notion of infalling boundary conditions in the bulk, and we arrived at an apparently consistent picture in which scalar fluctuations sense the presence of the Misner string, are anyonic, and lead to dissipative quasi-normal modes. 

Clearly it would be of interest to study numerically these quasi-normal modes as well as the analogous problem of graviton fluctuations, and thus probe the structure of correlation functions of the boundary stress-energy tensor and other local operators. We hope to return to such studies in the future. 

The structure of the solutions involves several interesting features, and it may be of interest to ask how one might take the limit $n\to 0$ to match onto fluctuations in the Schwarzchild black hole. Of course, some features of TNAdS$_4$ black holes are smooth in the limit. We believe though that this limit is generally singular as there is no way to smoothly remove a physical Misner string singularity. This is in keeping with the `instantonic' interpretation of NUT charge.

Finally we note that in Appendix B, we have made some preliminary remarks about a possible inner product on the space of solutions. Perhaps the most interesting feature of this discussion, which deserves further scrutiny, is how one should treat singularities in the solutions that occur as $u\to 1$. Since the geometry possesses closed timelike curves beyond a Killing horizon at a finite value of $u$, perhaps the proper treatment would involve imposing suitable boundary conditions there. It would be interesting to give this a physical interpretation.

\section*{Acknowledgements}

The work of RGL was supported by the U.S. Department of Energy under contract DE-SC0015655. RGL also thanks the Perimeter Institute for support, where part of this research was carried out. We thank Luca Ciambelli, Nick Halmagyi, Kostas Kokkotas and Marios Petropoulos for discussions. The work of ACP was supported by the Hellenic Foundation for Research and Innovation (H.F.R.I.) under the ``First Call for H.F.R.I. Research Projects to support Faculty members and Researchers and the procurement of high-cost research equipment grant" (MIS 1524, Project Number: 96048).

\appendix
\section{Review of Taub-NUT-AdS$_4$ spacetimes}
\renewcommand{\theequation}{\thesection.\arabic{equation}}
\setcounter{equation}{0}

We present here a brief review of TNAdS$_4$ spacetimes following \cite{Leigh:2011au,Leigh:2012jv} (see also \cite{Chamblin:1998pz,Astefanesei:2004kn}). A metric generalizing (\ref{TNAdS4})  is given by
\begin{align}
\label{TNAdS4gen}
ds^2 = \frac{dr^2}{V_\kappa(r)}+(r^2+n^2)[d\theta^2+g_\kappa^2(\theta)d\phi^2]
-V_\kappa(r)[dt+4ng_\kappa^2(\theta/2)d\phi]^2\,,\end{align}
where
\beqn
V_\kappa(r)=\kappa\frac{(r^2-n^2)}{r^2+n^2}+\frac{-2Mr+\frac{1}{L^2}(r^4+6n^2r^2-3n^4)}{r^2+n^2}\,,
\eeqn
with
\begin{align*}
g_\kappa(\theta)=\begin{cases} \sin\theta &,\quad \kappa=1 \\
\theta &,\quad \kappa=0 \\
\sinh\theta &,\quad \kappa=-1 \end{cases}\,.
\end{align*}
The $\kappa=1,0,-1$ cases distinguish the so-called spherical, planar or hyperbolic horizons. The metric is defined over $\theta<\pi$ for $\kappa=1$ and $\theta\in\mathbb{R}$ for $\kappa=-1,0$. {In the spherical case, the spacetime contains a Misner string singularity. This emanates from the fixed point (NUT) at $\theta=\pi$ (we assume $n>0$ throughout) of the Killing vector $\pa_\phi$ at the outer horizon $r_+$. The Misner string extends to $r=\infty$. 

One way to understand the string singularity is to note that in the natural choice of co-frame,
\beq
\hat e^0_N=\sqrt{V_1(r)}(dt+2n(1-\cos\theta)d\phi),\quad
\hat e^1_N=\sqrt{r^2+n^2}d\theta,\quad
\hat e^2_N=\sqrt{r^2+n^2}\sin\theta d\phi,\quad
\hat e^3_N=\frac{dr}{\sqrt{V_1(r)}}\,,
\eeq
the form $\hat e^0$ is ill-defined at $\theta\to\pi$ since $\phi$ is compact and $g_1^2(\theta/2)=\sin^2(\theta/2)\to 1$ rather than zero. A co-frame valid on the patch $\theta>0$ has $\hat e^0_S=\sqrt{V_1(r)}(dt-2n(1+\cos\theta)d\phi)$. These two choices of co-frame differ by $\hat e^0_N-\hat e^0_S= 4n\sqrt{V_1(r)}d\phi$ on the domain $0<\theta<\pi$. The induced co-frame on the asymptotic boundary similarly satisfies $\hat e^0_N-\hat e^0_S= \frac{4n}{L}d\phi$, that is they differ by a `gauge transformation' on the overlap of the coordinate patches. We thus have
\beq
\oint \hat e^0_N-\oint \hat e^0_S=\frac{8\pi n}{L}\,,
\eeq
which coincides with the total circulation. This is the invariant way to express the quantization condition corresponding to the compactification of the time direction; otherwise, the Misner string is physical and represents a point source of torsion. 

The thermodynamics of Taub-NUT AdS spacetimes is typically studied by analytically continuing $t\mapsto i\tau$ and $n\mapsto i \nu$ such that the Hawking temperature is given by 
\be
\label{TNtemp}
T_H=\frac{V'(r_+)}{4\pi}= \frac{L^2+3r_+^2-3\nu^2}{4\pi L^2 r_+}\to\frac{L^2+3r_+^2+3n^2}{4\pi L^2 r_+}
\ee
In recent papers \cite{Kubiznak:2019yiu,Bordo:2019tyh}, it has been noted that if one does not require the compactness of the time direction to be determined by $8\pi n/L$, then the NUT charge is freed up to play the role of a thermodynamical variable. Given the close similarity to the physics of magnetic fields, it is indeed natural to think of the NUT charge as the analogue of the magnetic field and its thermodynamic dual variable would be a 'magnetization'.  

\bigskip

The  isometry of (\ref{TNAdS4gen}) is generated by the Killing vectors  
\beqn
\label{TNAdS4Kill}
{\boldsymbol \xi}_1 &=& -\sin\phi \frac{g'_\kappa(\theta)}{g_\kappa(\theta)}\,\partial_\phi +\cos\phi\,\partial_\theta -4n\sin\phi\frac{g_\kappa(\theta/2)^2}{g_\kappa(\theta)}\,\partial_t\,, \\
{\boldsymbol \xi}_2 &=& \cos\phi \frac{g'_\kappa(\theta)}{g_\kappa(\theta)}\,\partial_\phi + \sin\phi\,\partial_\theta +4n\cos\phi\frac{g_\kappa(\theta/2)^2}{g_\kappa(\theta)}\,\partial_t\,, \\
{\boldsymbol \xi}_3 &=& \partial_\phi -2\kappa n\partial_t, \qquad 
{\bf e} = \partial_t\,.
\eeqn
with Lie brackets 
\beq
\left[{\boldsymbol \xi}_1,{\boldsymbol \xi}_2\right]=-\kappa{\boldsymbol \xi}_3,\qquad
\left[{\boldsymbol \xi}_3,{\boldsymbol \xi}_1\right]=-{\boldsymbol \xi}_2,\qquad
\left[{\boldsymbol \xi}_2,{\boldsymbol \xi}_3\right]=-{\boldsymbol \xi}_1,\qquad
 \left[{\boldsymbol \xi}_i,{\bf e}\right]=0.
\eeq
Thus for $\kappa=1$, we have the $SU(2)\times \mathbb{R}$ algebra and for $\kappa=-1$, we have $SL(2,\mathbb{R})\times\mathbb{R}$ isometry.
The form of the generator ${\boldsymbol \xi}_3$ is the source of most of the intrigue of this paper.
For the rest of the appendix, we will proceed with the spherical case, $\kappa=1$. Setting $u=\sin^2(\theta/2)$, 
the domain $\theta <\pi$ is mapped to the unit disk, $u<1$, and
\beqn
\label{TNAdS4Killu}
{\boldsymbol \xi}_1 &=& -\sin\phi \frac{1-2 u}{2\sqrt{u(1- u)}}\,\partial_\phi +\cos\phi\,\sqrt{u(1- u)}\partial_u -2n\sin\phi\sqrt{\frac{u}{1- u}}\,\partial_t\,, \\
{\boldsymbol \xi}_2 &=& \cos\phi \frac{1-2 u}{2\sqrt{u(1- u)}}\,\partial_\phi + \sin\phi\,\sqrt{u(1- u)}\partial_u +2n\cos\phi\sqrt{\frac{u}{1- u}}\,\partial_t\,, \\
{\boldsymbol \xi}_3 &=& \partial_\phi -2 n\partial_t, \qquad 
{\bf e} = \partial_t\,.
\eeqn
We note that these vector fields are not metrically positive definite (as a function of $u$, for all $n>0$), and this fact will have an important impact on the scalar solutions.
A usual trick in studying the representation theory of this algebra is to first {\it complexify} by defining 
${\bf L}_3=-i{\boldsymbol \xi}_3=-i(\partial_\phi-2 n\partial_t)$, ${\bf L}_\pm =\pm{\boldsymbol \xi}_1+ i {\boldsymbol \xi}_2 $, whereby
\begin{align}
 {\bf L}_\pm = \frac{ie^{\pm i\phi}}{\sqrt{u(1- u)}}&\left[2nu\partial_t \mp iu(1- u)\partial_u +\frac{1-2 u}{2}\partial_\phi\right] \,,
\end{align}
which satisfy  
\begin{align}
[ {\bf L}_+, {\bf L}_-]=2  {\bf L}_3 \qquad [ {\bf L}_3, {\bf L}_\pm] = \pm  {\bf L}_\pm
\end{align}
The quadratic $SU(2)$ Casimir is given by (see (\ref{L2}))
\begin{align}
{\bf L}^2 = 
-\sum_i{\boldsymbol \xi}_i^2= -\left\{ \partial_u[u(1- u)\partial_u] + \frac{4 n^2}{1- u}\partial_t^2 -\frac{2n}{1- u}\partial_t\partial_\phi +\frac{1}{4u(1- u)}\partial_\phi^2 \right\}\,.
\end{align}
In typical quantum mechanical applications, this complexification amounts to a formal trick, because we seek representations of the algebra on a complex vector space for which the generators are self-adjoint. It is well-known that if this is possible, one attains a unitary representation which in the case of $SU(2)$ is finite dimensional. The self-adjointness of the generators is not however automatic, as we are representing them on function spaces, and we have a non-compact algebra, $SU(2)\times\mathbb{R}$. 
Let us review some of the details of this issue, as it will be important for the proper treatment of TaubNUT.

Indeed, in the present case, we are not doing quantum mechanics, and are merely solving real differential equations on a real function space, under a choice of physically motivated boundary conditions.  
For any field fluctuation on the TNAdS$_4$ background, we can separate solutions possessing definite values for the quadratic Casimir. Here we study only scalar fluctuations, and as discussed in the body of the paper, we diagonalize the action of ${\bf L}^2$, $ {\bf L}_3$ and ${\bf e}$.
\beq
{\bf L}^2\Phi(u,\phi,t)=C\Phi(u,\phi,t),\qquad
{\bf L}_3\Phi(u,\phi,t)=m\Phi(u,\phi,t),\qquad
i\pa_t\Phi(u,\phi,t)=\omega\Phi(u,\phi,t)\,.
\eeq
Of course, we cannot actually diagonalize the two first order differential operators on real functions, but we sidestep that issue, as usual, by writing
\beq\label{diagsols}
\Phi(u,\phi,t)=\Phi_{\omega,q,m}(u)e^{-i\omega t}e^{i{\cal N}\phi},\qquad
C=q(q+1),\qquad {\cal N}=m-2n\omega\,,
\eeq
 and we write $\Omega=2n\omega$. The unusual $\phi$-dependence is a result of the form of the ${\bf L_3}$ generator. Note that $\hat{\bf L}^2$ is a real (but not positive) operator.

Let us now consider the Casimir equation ${\bf L}^2\Phi(u,\phi,t)=C\Phi(u,\phi,t)$ in detail, which in fact can be cast as a hypergeometric differential equation.
Solutions scale as $u^{\pm(m-\Omega)/2}$ near $u\sim0$, and as $(1-u)^{\pm(m+\Omega)/2}$ near $u\sim1$, and can be written in the form 
\beq
\Phi_{{\cal N},q,m}(u)=u^{a/2}(1-u)^{b/2} {}_2F_1(1+q+\tfrac{a+b}{2},-q+\tfrac{a+b}{2},1+a;u)
\eeq
where $a=\pm{\cal N}, b=\pm(2m-{\cal N})$. The solutions are not all independent of course.

Generally, we may construct highest/lowest weight representations (hwr/lwr) by constructing a section that is annihilated by ${\bf L}_\pm$. These are first-order differential equations that can be studied using standard properties of hypergeometrics. 
First we note that for functions of the form \eqref{diagsols}, we have
\begin{align}
{\bf L}_\pm = \frac{\pm e^{\pm i\phi}}{\sqrt{u(1- u)}}\left[  u(1- u)\partial_u \mp\frac12{\cal N}\pm mu\right]
\end{align}
The solutions given above are
\beqn
\Phi^{++}_{{\cal N},q,m}(u)&=&
u^{{\cal N}/2}(1-u)^{(2m-{\cal N})/2}{}_2F_1(1+q+m,-q+m,1+{\cal N};u),
\\
\Phi^{--}_{{\cal N},q,m}(u)&=&
u^{-{\cal N}/2}(1-u)^{-(2m-{\cal N})/2}{}_2F_1(1+q-m,-q-m,1-{\cal N};u),
\\
\Phi^{+-}_{{\cal N},q,m}(u)&=&
u^{{\cal N}/2}(1-u)^{-(2m-{\cal N})/2}{}_2F_1(1+q+{\cal N}-m,-q+{\cal N}-m,1+{\cal N};u),
\\
\Phi^{-+}_{{\cal N},q,m}(u)&=&
u^{-{\cal N}/2}(1-u)^{(2m-{\cal N})/2}{}_2F_1(1+q+m-{\cal N},-q+m-{\cal N},1-{\cal N};u).
\eeqn
In fact, by contiguous relations $\Phi^{+-}=\Phi^{++}$ and $\Phi^{-+}=\Phi^{--}$, so we can discard the third and fourth, and write $\Phi^+=\Phi^{++}, \Phi^-=\Phi^{--}$.
$\Phi^+$ is regular at $u\sim 0$ for ${\cal N}>0$ while $\Phi^-$ at $u\sim 0$ is regular for ${\cal N}<0$.  By direct computation, we find\footnote{The $SU(2)$ properties of these hypergeometrics follow from the differential relations
\beqn
\Big[x(1-x)\frac{d}{dx}+\gamma-1-x(\alpha+\beta-1)\Big]F(\alpha,\beta,\gamma;x)&=&
(\gamma-1)F(\alpha-1,\beta-1,\gamma-1;x)
\\
\Big[x(1-x)\frac{d}{dx}+(\gamma-1)(1-x)\Big]F(\alpha,\beta,\gamma;x)
&=&(\gamma-1)(1-x) F(\alpha,\beta,\gamma-1;x)
\\
\Big[x(1-x)\frac{d}{dx}-(\alpha+\beta-\gamma)x\Big]F(\alpha,\beta,\gamma;x)
&=&\frac{x}{\gamma}(\gamma-\alpha)(\gamma-\beta)F(\alpha,\beta,\gamma+1;x)
\\
\frac{d}{dx}F(\alpha,\beta,\gamma;x)
&=&\frac{\alpha\beta}{\gamma}F(\alpha+1,\beta+1,\gamma+1;x)
\eeqn
}
\beqn
{\bf L}_-\Phi^{+}_{{\cal N},q,m}(u)
&=&
-{\cal N}\Phi^{+}_{{\cal N}-1,q,m-1}(u) \,,
\\
{\bf L}_-\Phi^{-}_{{\cal N},q,m}(u)\,,
&=&
\frac{(1+q-m)(q+m)}{1-{\cal N}}
\Phi^{-}_{{\cal N}-1,q,m-1}(u)\,,
\label{Lmlws}
\\
{\bf L}_+\Phi^{+}_{{\cal N},q,m}(u)
&=&
\frac{(1+q+m)(-q+m)}{1+{\cal N}}\Phi^{+}_{{\cal N}+1,q,m+1}(u)
\label{Lphws}
\\
{\bf L}_+\Phi^{-}_{{\cal N},q,m}(u)\,,
&=&
-{\cal N}\Phi^{-}_{{\cal N}+1,q,m+1}(u)\,.
\eeqn
Thus we see that the solutions will indeed form $SU(2)$ representations.

Before proceeding, we can derive from the above
\beqn
{\bf L}_-{\bf L}_+\Phi^{+}_{{\cal N},q,m}(u)
&=&
(q+m+1)(m-q)\Phi^{+}_{{\cal N},q,m}(u)
\\
&=&
\Big(q(q+1)-m(m+1)\Big)\Phi^{+}_{{\cal N},q,m}(u)\,,
\\
{\bf L}_+{\bf L}_-\Phi^{+}_{{\cal N},q,m}(u)
&=&(q+m)(q-m+1)\Phi^{+}_{{\cal N},q,m}(u)
\\
&=&\Big(q(q+1)-m(m-1)\Big)\Phi^{+}_{{\cal N},q,m}(u)\,,
\\
{\bf L}_-{\bf L}_+\Phi^{-}_{{\cal N},q,m}(u)
&=&
(q+m+1)(q-m)\Phi^{-}_{{\cal N},q,m}(u)
\\
&=&
\Big(q(q+1)-m(m+1)\Big)\Phi^{-}_{{\cal N},q,m}(u)\,,
\\
{\bf L}_+{\bf L}_-\Phi^{-}_{{\cal N},q,m}(u)
&=&(q+m)(q-m+1)\Phi^{-}_{{\cal N},q,m}(u)
\\
&=&\Big(q(q+1)-m(m-1)\Big)\Phi^{-}_{{\cal N},q,m}(u)\,,
\eeqn
and putting these together, we find consistency with the Casimir, that is
\beqn
{\bf L}^2\Phi^{\pm}_{{\cal N},q,m}(u)
=
\Big({\bf L}_3^2+\tfrac12{\bf L}_+{\bf L}_-+\tfrac12{\bf L}_-{\bf L}_+\Big)\Phi^{\pm}_{{\cal N},q,m}(u)
=q(q+1)\Phi^{\pm}_{{\cal N},q,m}(u)\,.
\eeqn
Further details of the representations were given in the body of the paper and will not be repeated here. 

Note that we have arranged for the hws and lws to be regular at $u\sim 0$. There are several important comments to be made. First, it should be clear that $SU(2)$ descendants will eventually diverge at the origin; this behaviour seems reminiscent of phenomena in near-extremal black holes known as the Aretakis instability \cite{Aretakis:2012bm,Aretakis:2012ei}. See Refs. \cite{Hadar:2017ven,Hadar:2018izi} for recent discussions. 

\section{Inner Product}
\setcounter{equation}{0}

We have constructed a consistent picture of dissipative modes on AdSTN$_4$. In the body of the paper we found
\beqn
Y_{q,m,\Omega}(u,\phi)&=&
e^{i(m-\Omega)\phi} u^{(m-\Omega)/2}(1-u)^{(m+\Omega)/2} {}_2F_1(1+q+m,-q+m,1+m-\Omega;u)\,,
\\
(Y_{q,m,\Omega}(u,\phi)^*&=&
e^{-i(m^*-\Omega^*)\phi} u^{(m^*-\Omega^*)/2}(1-u)^{(m^*+\Omega^*)/2} {}_2F_1(1+q^*+m^*,-q^*+m^*,1+m^*-\Omega^*;u)\,.\nonumber
\eeqn
and in the case $q=-\frac12-i\Omega_2$, these are $SU(2)$ dual representations with real Casimir. 
Given that these functions depend on both $m\pm\Omega$, it is natural to introduce the vector fields 
\beq
{\bf K}_\pm= {\bf L}_3\mp 2in\,{\bf e}
\eeq
We see that
\beq
{\bf K}_-=-i\pa_\phi,\qquad {\bf K}_+=-i(\pa_\phi-4n\,\pa_t),\qquad {\bf L}_3=\frac12({\bf K}_++{\bf K}_-)
\eeq
The vector field ${\bf K}_+$ has been claimed to be relevant to the thermodynamics of AdSTN$_4$ (see 
\cite{Durka:2019ajz,Bordo:2019tyh,Kubiznak:2019yiu} and \cite{Carlip:1999cy}). 

One of the properties that we have imposed throughout the paper is the reality of the eigenvalues of ${\bf K}_-$. This would seem to imply that ${\bf K}_-$ should be thought of as corresponding to a self-adjoint operator on the space of solutions. To make this more precise, we would need to introduce an inner product on the space of solutions. Given that the solutions described in the text fall into highest- (${\cal R}$) and lowest-weight (${\cal R}^*$) representations, it is natural to introduce the $SU(2)$-invariant product ${\cal R}\times {\cal R}^*$  which reads
\beq
\langle \bar q,\bar m,\bar\Omega|q,m,\Omega\rangle
=2\int du \int_0^{2\pi} d\phi\ Y_{\bar q',\bar m',\bar\Omega'}(u,\phi) Y_{q,m,\Omega}(u,\phi).
\eeq
 To be non-zero, we must have $\bar q'=q^*$, $\bar m'=-m^*$ and $\bar\Omega'=-\Omega^*$.  The adjoint $\hat {\cal O}^\dagger$ of an operator $\hat {\cal O}$ satisfies 
\beq
2\int du \int_0^{2\pi} d\phi\ Y_{\bar q,\bar m,\bar\Omega}(u,\phi) (\hat{\cal O}Y_{q,m,\Omega}(u,\phi))
=2\int du \int_0^{2\pi} d\phi\ (\hat{\cal O}^\dagger Y_{\bar q,\bar m,\bar\Omega}(u,\phi)) Y_{q,m,\Omega}(u,\phi)\,.
\eeq
As usual, the only subtlety in having $\hat{\cal O}$ self-adjoint is that the two sides differ by an integration by parts. In the case of ${\bf K}_-$, this is the condition 
\beq
\left.\left[Y_{\bar q,\bar m,\bar\Omega}(u,\phi) Y_{q,m,\Omega}(u,\phi)\right]\right|^{2\pi}_0=0
\eeq
which is of course satisfied (even though $Y_{q,m,\Omega}(u,\phi)$ is not itself periodic) for $m-\Omega\in\mathbb{R}$. 
The operator ${\bf K}_+$ is not self-adjoint. There is a further subtlety with ${\bf K}_-$, regarded as a vector field at the conformal boundary of TNAdS$_4$: it is space-like only on the domain $u\in (0,u_*)$, where $u_*=\frac{1}{1+4n^2/L^2}$. This of course is the place at which $g_{\phi\phi}$ passes through zero, and in these coordinates, there are thus closed time-like curves beyond. Perhaps we should take this as an indication that in the definition of the inner product, the integration over $u$ should extend not over $u\in [0,1]$, but instead over $u\in (0,u_*)$. This in fact would be helpful, as the integrand blows up uncontrollably for $u\to 1$, and hence any integrals on $u\in [0,1]$ would not exist. Perhaps ultimately what we should do is to treat the system as having a horizon at fixed $u$ and introduce appropriate boundary conditions. We leave this for future analysis.

\myfig{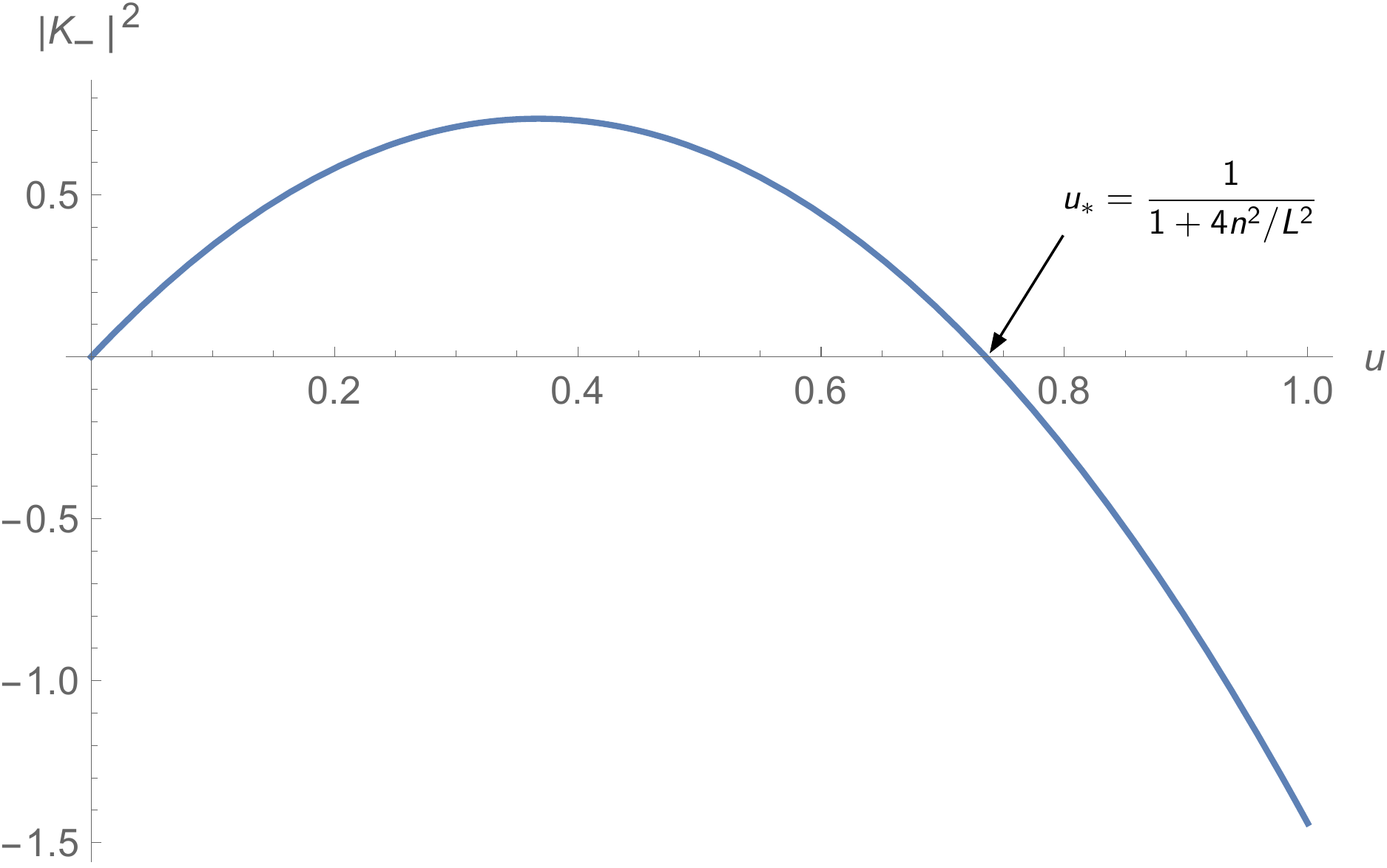}{6}{The norm of the vector ${\bf K}_-$. There are horizons at $u=0, \frac{1}{1+4n^2/L^2}$.}

\providecommand{\href}[2]{#2}\begingroup\raggedright\endgroup
\end{document}